\documentclass[preprint,amsmath,aps,10pt,superscriptaddress,letterpaper,tightenlines,showpacs]{revtex4}
\usepackage{bm}
\usepackage{stmaryrd}
\usepackage{amsmath}
\usepackage{amssymb}
\usepackage{graphicx}
\usepackage{textcomp}
\usepackage{calrsfs}
\usepackage{yfonts}
\usepackage{color}
\newcommand{\Xb}{\textbf{X}}

\newcommand{\rb}{\textbf{r}}
\newcommand{\Fb}{\textbf{F}}
\newcommand{\Gb}{\textbf{G}}
\newcommand{\Jb}{\textbf{J}}
\newcommand{\Xbar}{\bar{\boldsymbol{\chi}}}
\newcommand{\ob}{\hat{\textbf{o}}}

\newcommand{\gb}{\textbf{g}}

\newcommand{\p}{{\partial}}

\newcommand{\haf}{{\frac{1}{2}}}
\newcommand{\intf}{{\int\limits_0^{\infty}\,}}

\newcommand{\ra}{{\rangle}}
\begin{document}

\title{From Brownian motion formalism to fluctuation-induced force in a general fluctuating medium}

\author{Fardin Kheirandish}

\affiliation{Department of Physics, Faculty of Science, University of Kurdistan, P.O.Box 66177-15175, Sanandaj, Iran}

\begin{abstract}
Starting from a microscopic approach and using the formalism of quantum Brownian motion, partition function of a system composed of two separated pieces of anisotropic matter and a fluctuating medium in finite temperature is obtained rigorously. A general expression for fluctuation-induced free energy between the separated anisotropic pieces of matter is obtained and it is shown that in the framework of induced-force, the free energy of mean-force and effective free energy are equivalent.
\end{abstract}
\maketitle

\section{Introduction}
\noindent Fluctuation-induced forces are ubiquitous phenomena in a wide variety of systems in physics and chemistry \cite{Moste, Krech, Weinberg}. Since the seminal paper of Casimir \cite{Casimir1948} on fluctuation induced force between
two parallel plates made of perfect conductors due to vacuum fluctuations of
electromagnetic field and its generalization to the case of dielectric slabs by Lifshitz \cite{Lifshitz1956, Lifshitz1961}, an extensive work has
been down on fluctuation-induced forces \cite{Israel1992, Kardar1999, Bordag2001, Milton2001, Bordag2008, Kh-1, Kh-2, Kh-3, Kh-4, Marjan1}.
Alongside the theoretical works, experimental high precision verifications of the Casimir force have been achieved
\cite{Lamoreaux97, Lamoreaux98, Lamoreaux99, MR98, Harris, CAKBC2001, Bressi02}.

Electromagnetic based fluctuation-induced forces Known as Casimir forces, impose serious constraints on mechanical motions of nanoscale parts of a nano-machine were Casimir forces cause considerable friction leading to stiction \cite{Serry1995, Serry1998, Buks2001, Buks2002, Srivastava+85, Stroscio+86}.

In thermal equilibrium and in the framework of quantum Hamiltonian of mean force (QHMF) \cite{Campisi, Hilt, Marjan2}, the equilibrium free energy of a subsystem in contact with its environment is equal to the difference between the free energy of the total system and free energy of the solely its environment \cite{Campisi}. The key to overcome the calculation of this quantity is the QHMF. The QHMF is the effective Hamiltonian that describes the Boltezmann-Gibbs equilibrium of the probability density of the open quantum subsystem of interest.

In the present letter, motivated by the Euclidean partition function approach to calculate the Casimir energy between separated objects in electromagnetic quantum vacuum, using the formalism of quantum Brownian motion of a collection of quantum harmonic oscillators in a fluctuating field, we first rigorously find a general expression for free energy of two separated anisotropic pieces of matter interacting linearly with a general fluctuating medium in finite temperature. This expression is important from both theoretical and numerical point of view. Then we will show that as far as the fluctuation-induced force between separated material objects is considered, the free energy of mean force and the effective free energy obtained from the interacting part of the total partition function, are equivalent.
\section{Model}
\noindent Let $A_1$ and $A_2$ be two separate pieces of anisotropic matter interacting linearly with a general fluctuating field $\Fb(\rb,t)$ trough coupling tensors $g^{(1)}_{ij}$ and $g^{(2)}_{ij}\,\,\,(i,j=1,2,3)$, see Fig.1. We take $A_1 +A_2$ as our main subsystem and want to find an effective free energy for this subsystem by making use of Euclidean path integrals and tracing out the environmental degrees of freedom.  
\begin{figure}
\includegraphics[width=3in]{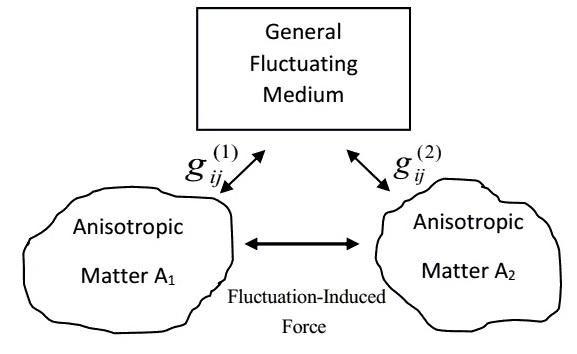}
\caption{A general fluctuating medium interacting linearly with to pieces of anisotropic matter.}
\label{fig.1}
\end{figure}
The total Lagrangian density can be described by
\begin{eqnarray}\label{L}
\mathcal{L} &=& -\haf \Fb\cdot\ob\cdot\Fb-\haf\intf d\nu\,\Xb_\nu\cdot(\p_t^2+\nu^2)\,\Xb_\nu\nonumber\\
&+& \intf d\nu\,\Fb\cdot\bar{\gb}\cdot\p_t\Xb_\nu,
\end{eqnarray}
where for notational simplicity we have assumed
\begin{eqnarray}\label{simplicity}
  \Fb\cdot\ob\cdot\Fb &=& \sum_{i,j=1}^3 F_i\,\hat{o}_{ij}\,F_j,\nonumber\\
  \Fb\cdot\bar{\gb}\cdot\p_t\Xb_\nu &=& \sum_{i,j=1}^3 F_i\,g_{ij}(\nu,\rb)\,\p_t X_{\nu,j}.
\end{eqnarray}
The first term in Eq.(\ref{L}) describes the Lagrangian density of the fluctuating field and the second therm is the Lagrangian density of the anisotropic matter. Note that, as is usual in literature, the matter fields are described by a continuum of harmonic oscillators with position vector field $\Xb_\nu$.  The operator $\ob$ can be a general linear operator. For example, when the fluctuating field is the electromagnetic field, then $\ob=\p_t^2+\nabla\times\nabla\times$, \cite{Kh-4, Marjan1}. The fluctuating field can be assumed as a scalar, vector, tensor or a spinor field interacting linearly with material fields and the only modification in each case is a rearrangement of indices on fields and coupling functions in the total Lagrangian density. Here we have considered a vector fluctuating field. Note that the anisotropic matter is modeled by a continuum of Lagrangian densities of quantum harmonic oscillators \cite{Hopfield, Huttner, Huttner+92, Barnett, Kh-5, Kh-5-2, Kh-6, Kh-7, Kh-8}. The third term in total Lagrangian density describes the linear interaction between anisotropic matter and fluctuating field. The strength of the coupling is described by the coupling tensor $\bar{\gb} (\nu,\rb)$ which is nonzero inside the regions $A_1$ and $A_2$ and zero outside these regions. The coupling tensor mathematically is described by
\begin{equation}\label{F}
\gb_{ij}(\nu,\rb)=\left\{
  \begin{array}{ll}
    g^{(1)}_{ij} (\nu), &  \rb\in A_1; \\
    g^{(2)}_{ij} (\nu), &  \rb\in A_2; \\
    0, & \mbox{otherwise}.
  \end{array}
\right.
\end{equation}
From Heisenberg equations of motion, we find the following equations for fluctuating and material fields, respectively
\begin{eqnarray}
 && \ob\cdot\Fb (\rb,t)= \intf d\nu\, \bar{\gb}(\nu,\rb)\cdot\p_t\Xb_\nu (\rb,t),\label{Eqf}\\
 && (\p_t^2+\nu^2)\,\Xb_\nu (\rb,t) = -\bar{\gb}(\nu,\rb)\cdot\p_t\Fb (\rb,t). \label{Eqx}
\end{eqnarray}
Equation (\ref{Eqx}) can be solved formally as
\begin{equation}\label{Eqx2}
  \Xb_\nu (\rb,t)=\Xb_\nu^{(n)} (\rb,t)-\int_0^t dt'\,G_\nu (t-t')\bar{\gb}\cdot\p_{t'}\Fb (\rb,t'),
\end{equation}
where $G_\nu (t-t')$ is the retarded Green's function that can be expressed in terms of Heaviside step function as
\begin{equation}\label{Green}
 G_\nu (t-t')=\Theta(t-t')\,\frac{\sin\nu (t-t')}{\nu},
\end{equation}
and $\Xb_\nu^{(n)}$, is the homogeneous solution $(\p_t^2+\nu^2)\,\Xb_\nu^{(n)}=0$, or material quantum noise field. By inserting the solution (\ref{Eqx2}) into (\ref{Eqf}), we find the quantum Langevin equation for the fluctuating field in the presence of material fields
\begin{eqnarray}\label{FE}
  && \ob\cdot\Fb (\rb,t)+\p_t \int_0^t dt'\,\Xbar(\rb,t-t')\cdot\p_{t'} \Fb (\rb,t')\nonumber\\
  && =\intf d\nu\,\bar{\gb}\cdot\p_t \Xb_{\nu}^{(n)}(\rb,t),
\end{eqnarray}
where the response or memory tensor is defined by
\begin{equation}\label{kapa}
 \Xbar (\rb,t-t')=\intf d\nu\, \bar{\gb}\cdot\bar{\gb}\,G_\nu (t-t').
\end{equation}
For notational simplicity, in Eq.(\ref{kapa}) we have assumed that the coupling tensors are symmetric $\bar{\gb}=\bar{\gb}^t$ \cite{Kh-8}, that is the imaginary part of the susceptibility tensor is symmetric, see Eq.(\ref{relation}). One can proceed without this assumption and consider
$\bar{\gb}\cdot{\bar{\gb}}^t$ instead of $\bar{\gb}\cdot\bar{\gb}$.  Equation (\ref{kapa}) is a sine transform and its inverse leads to the following relation between coupling and memory tensor in frequency space
\begin{equation}\label{relation}
  \bar{\gb}(\nu,\rb)=\sqrt{\frac{2\nu}{\pi}\,\mbox{Im}[\Xbar(\rb,\nu)]}.
\end{equation}
Therefore, if we are given a definite response tensor, we can adjust the coupling tensor according to Eq.(\ref{relation}).
\section{Partition function}
\noindent To find the partition function of the subsystem described by material pieces, we first switch to Euclidean Lagrangian $\mathcal{L}_E $, obtained by a Wick rotation on time coordinate, this implies
\begin{equation}\label{wick}
  \left\{
    \begin{array}{ll}
      i t=\tau \\
      \p_t=i\p_\tau
    \end{array}
  \right.
\Rightarrow\ob(\p_t^2,\p_i) \rightarrow \ob'(-\p_\tau^2,\p_i),
\end{equation}
and all fields are now functions of $(\rb,\tau)$. The total partition function is defined by \cite{Kh-4}
\begin{eqnarray}\label{Z}
Z &=& \int \prod_{\nu\geq 0} D[\Xb_\nu ]{D[\Fb]\,e^{-\haf\int d\rb\int_{0}^{\beta} d\tau [\Fb\cdot\ob'\cdot\Fb+\Fb\cdot{\Jb}]}}\nonumber\\
& \times & e^{-\haf\int d\rb\int_{0}^{\beta} d\tau\intf d\nu\,\Xb_\nu \cdot(-\p_\tau^2+\nu^2)\,\Xb_\nu},
\end{eqnarray}
where $\beta=1/k_B\,T$, $k_B$ is Boltzmann constant and $T$ is the temperature of the fluctuating medium described by the field $\Fb$. The source term $\Jb$ in Eq.(\ref{Z}) is defined by
\begin{equation}\label{J}
 \Jb (\rb,\tau)=i\intf d\nu\,\bar{\gb}\cdot\p_\tau \Xb_\nu (\rb,\tau).
\end{equation}
To find the partition function, periodic boundary conditions on Bosonic fields are imposed
\begin{eqnarray}\label{BC}
\Fb(\rb,\tau) &=& \Fb(\rb,\tau+\beta)\nonumber\\
&=& \sum_{n=0}^{\infty\,'} [\Fb_n (\rb) e^{-i\omega_n \tau}+c.c.],\nonumber\\
\Xb_\nu (\rb,\tau) &=& \Xb_\nu (\rb,\tau+\beta)\nonumber\\
&=& \sum_{n=0}^{\infty\,'} [\Xb_{\nu,n} (\rb) e^{-i\omega_n \tau}+c.c.],
\end{eqnarray}
where $\omega_n=2\pi n/\beta$ are Matsubara frequencies and the prime over the summation means
the term corresponding to $n = 0$, should be given half
weight. Inserting Eqs.(\ref{BC}) into Eq.(\ref{Z}), we find
\begin{eqnarray}
Z &=& \int \prod_{n,\nu\geq 0} D[\Xb_{\nu,n}]D[\Xb^{*}_{\nu,n}]\prod_{n\geq 0}D[\Fb_n]D[\Fb^{*}_n]\nonumber\\
&\times & e^{-\haf
\int d\rb\,\sum\limits_{n=0}^{\infty\,'} (\Fb_n\cdot\beta\ob_n\cdot\Fb^{*}_n +\Fb^{*}_n\cdot\beta\ob_n\cdot\Fb_n
+\Fb_n\cdot\Jb^{*}_n+\Fb^{*}_n\cdot\Jb_n)}\nonumber\\
& \times & e^{-\haf\int d\rb\intf d\nu\,(\Xb^{*}_{\nu,n}\cdot\beta(\omega_n^2+\nu^2)\,\Xb_{\nu,n}+
\Xb_{\nu,n}\cdot\beta(\omega_n^2+\nu^2)\,\Xb^{*}_{\nu,n})},\nonumber\\
\end{eqnarray}
where for convenience we have defined $\ob_n=\ob'(\omega_n^2,\p_i)$. By making use of the well known formula \cite{Zinn}
\begin{eqnarray}\label{formula}
 && \int D[\varphi]D[\varphi^*]\,e^{-\int d\rb\,(\varphi^* \hat{A}\varphi+\varphi \hat{A}\varphi^*+\rho\varphi^*+\varrho^*\varphi)}\nonumber\\
 && = (\det \hat{A})^{-1}\,e^{\int d\rb\, \rho^*\hat{A}^{-1} \rho},
\end{eqnarray}
we can integrate over fluctuating field and material degrees of freedom and find the total partition function as
\begin{eqnarray}\label{Zf}
Z &=&\underbrace{\prod_{n\geq 0}^{\infty\,'}(\det[\beta\ob_n])^{-1}}_{Z_F}\underbrace{\prod_{n\geq 0}^{\infty\,'}\prod_{\nu\geq 0}^{\infty}(\det[\beta(\omega_n^2+\nu^2)])^{-1}}_{Z_m}\nonumber\\
&\times & \underbrace{\prod_{n\geq 0}^{\infty\,'}\prod_{\nu\geq 0}^{\infty}(\det[1+\omega^2_{n}\,G_\nu (\omega_n)\,\bar{\gb}\cdot\textbf{G}_0\cdot\bar{\gb}])^{-1}}_{Z_{eff}},
\end{eqnarray}
where $\textbf{G}_0$ is the dyadic Green's function of the fluctuating field in free space $\ob_n\cdot\textbf{G}_0=\mathbb{I}$. In Eq.(\ref{Zf}), the first product term is the partition function of the fluctuating field ($Z_{F}$), the second product term is the partition function of the material field ($Z_{m}$) and the last term which is the relevant term for our purposes, originates from interaction between the fluctuating field and material field ($Z_{eff}$).
Using the identity $\ln[\det \hat{O}]=\mbox{Tr}\ln[\hat{O}]$, and definition of the relevant or effective free energy $F_{eff}=-k_B T \ln Z_{eff}$,  we find
\begin{equation}\label{Free1}
   F_{eff}=k_B T \sum_{n\geq 0}^{\infty\,\,'}\mbox{Tr}\ln[1+\omega^2_n\,G_\nu (\omega_n)\,\bar{\gb}\cdot\textbf{G}_0\cdot\bar{\gb}].
\end{equation}
By making use of the expansion
\begin{equation}\label{expan}
  \ln(1+x)=\sum_{m=1}^{\infty} (-1)^{m-1}\, \frac{x^{m}}{m},
\end{equation}
and Fourier transform of the memory or response tensor, Eq.(\ref{kapa})
\begin{equation}\label{suscepF}
\chi_{ij} (\rb,\omega) =\intf d\nu\,\frac{\gb_{ik}(\nu,\rb)\,\gb_{kj}(\nu,\rb)}{\omega^2+\nu^2},
\end{equation}
we find the free energy in terms of the response tensor as
\begin{eqnarray}\label{Free2}
 && F_{eff}=-k_B T\ln Z_{eff}\nonumber\\
 && =k_B T\sum_{n=0}^{\infty\,'} \mbox{Tr}_{|i,\rb\ra}\,
\ln[1+\omega_n^2\,\Xbar (i\omega_n)\cdot\bar{\Gb}_0 (i\omega_n)],
\end{eqnarray}
where $\mbox{Tr}_{|i,\rb\ra}$, means taking trace over position and internal degrees of freedom ($i=1,2,3$). The Green function of Eq.(\ref{FE}) satisfies
\begin{eqnarray}\label{MainGreen}
\ob_n\cdot\textbf{G}=\omega_n^2\,\Xbar\cdot\textbf{G}+\mathbb{I},
\end{eqnarray}
and by iteration we find
\begin{eqnarray}\label{MainGreen2}
\textbf{G}=\frac{1}{1-\omega_n^2\,\textbf{G}_0\cdot\Xbar}\cdot \textbf{G}_0,
\end{eqnarray}
therefore,
\begin{equation}\label{Elegant}
F_{eff}=-k_B T\,\ln(\textbf{G}\cdot \textbf{G}_0^{-1}).
\end{equation}
The equation (\ref{Elegant}) is known as the elegant formula \cite{Moste,Bordag2001,Milton2001,Bordag2008,Dalvit2011} in the literature, here this equation is derived for a general fluctuating field interacting linearly with anisotropic media in finite temperature. From computational point of view, one can find a series expansion in susceptibility tensor $\Xbar$. For this purpose let us expand the logarithm in (\ref{Free2}) using the expansion (\ref{expan}), we find
\begin{equation}\label{expansion}
  F_{eff}=k_B T\sum_{n=0}^{\infty\,'}\sum_{m=1}^{\infty}\frac{(-1)^{m-1}}{m}\, \mbox{Tr}_{|i,\rb\ra}\,(\Xbar (i\omega_n)\cdot\bar{\Gb}_0 (i\omega_n))^m,
\end{equation}
which is a generalization of the result reported in \cite{Ramin, Kh-4} for the case of electromagnetic field in the presence of isotropic matter.

In zero temperature, using the correspondence
\begin{equation}\label{zero}
  \intf \frac{d\zeta}{2\pi}\leftrightarrow k_B T\sum_{n=0}^{\infty\,'},
\end{equation}
we find
\begin{equation}\label{zero1}
  F=\intf \frac{d\zeta}{2\pi}\,\mbox{Tr}_{|i,\rb\ra}\,[\ln(1+\Xbar(i\zeta)\cdot\bar{\Gb}_0 (i\zeta))].
\end{equation}
\section{The induced-force and Hamiltonian of mean-force}
\noindent In this section, it is shown that the force induced between the separated pieced of matter due to the fluctuations of their medium can be equivalently calculated from the free energy of mean force of the material pieces. For this purpose, let the subsystem be separate pieces of matter defined by regions $A_1$ and $A_2$, interacting linearly with a fluctuating medium, Fig.1. The total Hamiltonian is
\begin{equation}\label{H}
  H=H_F+H_S+H_{int},
\end{equation}
where $H_F$ is the Hamiltonian of the fluctuating field, $H_S$ is the Hamiltonian of the subsystem ($A_1+A_2$), and $H_{int}$ is the interaction term. The reduced density matrix for the subsystem is defined by \cite{Campisi}
\begin{equation}\label{rhos}
  \rho_S =\frac{e^{-\beta H_S^*}}{Z^*},
\end{equation}
where $Z^*=\mbox{Tr} \exp(-\beta H_S^*)$, is the reduced partition function and $H_S^*$ is the Hamiltonian of mean force defined by \cite{Campisi}
\begin{equation}\label{HMF}
  H_S^*=-\frac{1}{\beta}\ln \frac{\mbox{Tr} \exp(-\beta H)}{\mbox{Tr} \exp(-\beta H_{F})}.
\end{equation}
From Eq.(\ref{HMF}) we have
\begin{equation}\label{HMF1}
  e^{-\beta H_S^*}=\frac{Z\,\rho_S}{Z_F},
\end{equation}
where $Z=\mbox{Tr}\exp(-\beta H)$ is the partition of the total system. Therefore,
\begin{equation}\label{Zstar}
  Z^*=\frac{Z}{Z_M}.
\end{equation}
The free energy of mean force is defined by
\begin{equation}\label{FMF}
 F^*=-K_B T\,\ln Z^*=-K_B T\,\ln Z/Z_F.
\end{equation}
From Eqs.(\ref{Zf},\ref{FMF}), we have
\begin{equation}\label{Final1}
  F^*=-K_B T\,(\ln Z_{eff}+\ln Z_m)=F_{eff}+F_m,
\end{equation}
where $F_m$ is the self energy of the material fields which is not our concern here. The fluctuation-induced force between material pieces is calculated from spatial derivative of free energy with respect to a relevant distance which appears in effective free energy $F_{eff}$, therefore, free energy of mean force and the effective free energy both lead to the same induced force between material pieces, that is $F^*\equiv F_{eff}$.
\section{conclusion}
\noindent Starting from a microscopic approach, a general expression for fluctuation-induced free energy
between two separate anisotropic pieces of matter was obtained in a general fluctuating medium in finite temperature. It was shown that, as far as the induced force between material pieces is considered, the free energy of mean-force
and the effective free energy are equivalent.


\begin{thebibliography}{0}
\bibitem{Moste} V. M. Mostepanenko and N. N. Trunov, \emph{The Casimir Effect and Its Applications} (Clarendon Press, Oxford, 1997).

\bibitem{Krech} M. Krech, \emph{The Casimir Effect in Critical Systems} (World Scientific, Singapore 1994).

\bibitem{Weinberg} S. Weinberg, Rev. Mod. Phys. \textbf{61}, 1 (1989).

\bibitem{Casimir1948} H. B. G. Casimir, Proc. K. Ned. Akad. Wet. \textbf{51}, 793 (1948).

\bibitem{Lifshitz1956} E. M. Lifshitz E. M.,Sov. Phys. JETP \textbf{2}, 73 (1956).

\bibitem{Lifshitz1961} I. E. Dzyaloshinskii, E. M. Lifshitz and L. P. Pitaevskii, Adv. Phys. \textbf{10}, 165 (1961).

\bibitem{Israel1992} J. N. Israelachvili, \emph{Intermolecular and Surface Forces} (Academic, London 1992).

\bibitem{Kardar1999} M. Kardar M. and R. Golestanian, Rev. Mod. Phys. \textbf{71}, 1233 (1999).

\bibitem{Bordag2001} M. Bordag, U. Mohideen and V. M. Mostepanenko, Phys. Rep. \textbf{353}, 1 (2001).

\bibitem{Milton2001} K. A. Milton, \emph{The Casimir Effect: Physical Manifestations of Zero-Point Energy}
(WorldSc ientific, Singapore 2001).

\bibitem{Bordag2008} M. Bordag, G. L. Klimchitskaya, U. Mohideen and V. M. Mostepanenko, \emph{Advances in the Casimir Effect}
(Oxford university press, 2008).

\bibitem{Dalvit2011} D. A. R. Dalvit, P. A. M. Neto, and F. D Mazzitelli,
\emph{Casimir Physics, Lecture Notes in Physics} (Springer-Verlag Berlin Heidelberg, 2011).

\bibitem{Kh-1} F. Kheirandish, E. Amooghorban and M. Soltani, Phys. Rev. A, \textbf{83}, 032507 (2011).

\bibitem{Kh-2} F. Kheirandish, M. Soltani M. and J. Sarabadani, Annals of Physics, \textbf{326}, 657 (2011).

\bibitem{Kh-3} E. Amooghorban, M. Wubs, A. Mortensen and. Kheirandish, Phys. Rev. A, \textbf{84}, 013806 (2011).

\bibitem{Kh-4} F. Kheirandish and S. Salimi, Phys. Rev. A, \textbf{84}, 062122 (2011).

\bibitem{Marjan1} F. Kheirandish and M. Jafari, Phys. Rev. A, \textbf{86}, 022503 (2012).

\bibitem{Lamoreaux97} S. K. Lamoreaux, Phys. Rev. Lett., \textbf{78}, 5 (1997).

\bibitem{Lamoreaux98} S. K. Lamoreaux, Phys. Rev. Lett., \textbf{81}, 5475 (1998).

\bibitem{Lamoreaux99} S. K. Lamoreaux, Phys. Rev. A, \textbf{59}, 3149 (1999).

\bibitem{MR98} U. Mohideen and A. Roy, Phys. Rev. Lett., \textbf{81}, 4549 (1998).

\bibitem{Harris} B. W. Harris B. W., F. Chen and U. Mohideen, Phys. Rev. A, \textbf{62}, 052109 (2000).

\bibitem{CAKBC2001} H. B. Chan, V. A. Aksyuk, R. N. Kleiman, D. J. Bishop  and F. Capasso, Science, \textbf{291}, 1941 (2001).

\bibitem{Bressi02} G. Bressi, G. Carugno, R Onofrio and G. Ruoso, Phys. Rev. Lett., \textbf{88}, 041804 (2002).

\bibitem{Serry1995} F. M. Serry, D. Walliser and  G. J. Maclay, J. Microelectromech. Syst., \textbf{4}, 193 (1995).

\bibitem{Serry1998} F. M. Serry, D. Walliser and G. J. Maclay, J. Appl. Phys., \textbf{84}, 2501 (1998).

\bibitem{Buks2001} E. Buks and M. L. Roukes, Phys. Rev. B, \textbf{63}, 033402 (2001).

\bibitem{Buks2002} E. Buks and M. L. Roukes, Nature \textbf{419}, 119 (2002).

\bibitem{Srivastava+85} Y. Srivastava, A Widom and M. H. Friedman, Phys. Rev. Lett., \textbf{55}, 2246 (1985).

\bibitem{Stroscio+86} M. A. Stroscio, Phys. Rev. Lett., \textbf{56}, 2107 (1986).

\bibitem{Campisi} M. Campisi, P. Talkner and P. Hanggi, Phys. Rev. Lett., \textbf{102}, 210401 (2009).

\bibitem{Hilt} S. Hilt, B. Thomas and E. Lutz, Phys. Rev. E, \textbf{84}, 031110 (2011).

\bibitem{Marjan2} M. Jafari and F. Kheirandish, Laser Phys., \textbf{27}, 015201 (2017).

\bibitem{Hopfield} J. J. Hopfield, Phys. Rev., \textbf{112}, 1555 (1958).

\bibitem{Huttner} B. Huttner and S. M. Barnett, Phys. Rev. A, \textbf{46}, 4306 (1992).

\bibitem{Huttner+92} B. Huttner and S. M. Barnett, Europhys. Lett., \textbf{18}, 487 (1992).

\bibitem{Barnett} S. Barnett, R. Matloob and R. Loudon, J. Mod. Opt., \textbf{42}, 1165 (1995).

\bibitem{Kh-5} F. Kheirandish and M. Amooshahi, Phys. Rev. A, \textbf{74}, 042102 (2006).

\bibitem{Kh-5-2} M. Amooshahi and F. Kheirandish, Phys. Rev. A, \textbf{76}, 062103 (2007).

\bibitem{Kh-6} F. Kheirandish, M. Amooshahi and M. Soltani, J. Phys. B: Mol. Opt. Phys., \textbf{42}, 075504 (2009).

\bibitem{Kh-7} F. Kheirandish and M. Soltani, Phys. Rev. A, \textbf{78}, 012102 (2008).

\bibitem{Kh-8} F. Kheirandish and M. Amooshahi, J. Phys. A: Math. Theor., \textbf{41}, 275402 (2008).

\bibitem{Zinn} J. Zinn-Justin, \emph{Quantum Field Theory and Critical Phenomena} (Oxford, Oxford University Press 1995).

\bibitem{Ramin} R. Golestanian, Phys. Rev. Lett., \textbf{95}, 230601 (2005).
\end{thebibliography}
\end{document}